# Data Equity:
# Foundational Concepts for Generative AI

BRIEFING PAPER

OCTOBER 2023

World Economic Forum

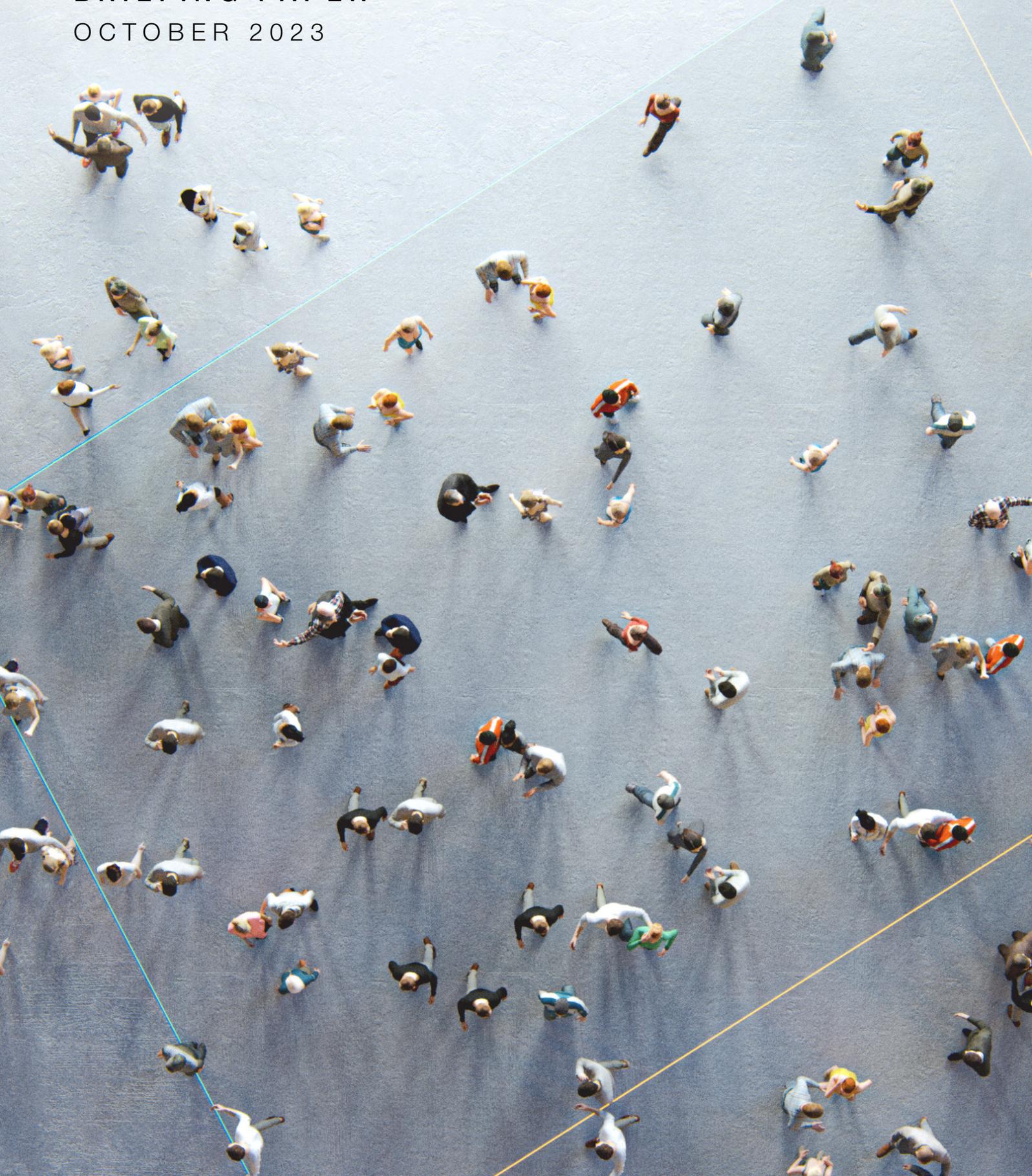

Images: Getty Images

# Contents

# Introduction

Over the past several months, a series of technological advances have emerged as a result of generative artificial intelligence (genAI) tools, including ChatGPT, Bard, Midjourney, and Stable Diffusion. The use of these tools has gained significant attention and captured the imagination of public and industry stakeholders due to its capabilities, wide range of applications and ease of use.

Given its potential to challenge established business practices and operational paradigms, and the promise of rapid innovation coupled with the likelihood of significant disruption, genAI is sparking global conversations. These anticipated, far-reaching consequences have a societal dimension and will require comprehensive engagement from key stakeholders such as industry, government, academia and civil society.

At the heart of these discussions lies the concept of "data equity" – a core notion within data governance centred on the impact of data on the equity of technical systems for individuals, groups, enterprises and ecosystems.[1] It includes concepts of data fairness, bias, access, control and accountability, all underpinned by principles of justice, non-discrimination, transparency and inclusive participation.

Data equity is not a new concept; it is grounded in human rights and part of ongoing work on data privacy, protection, ethics, Indigenous data sovereignty and responsibility. The intersection of data equity and genAI, however, is new and presents unique challenges. The datasets used to train AI models are prone to biases that reinforce existing inequities. This requires proactively auditing data and algorithms and intervening at every step of the AI process, from data collection to model training to implementation, to ensure that the resulting genAI tools fairly represent all communities. With the advent of genAI significantly increasing the rate at which AI is deployed and developed, exploring frameworks for data equity is more urgent than ever.

This briefing paper delves into these issues, with a particular focus on data equity within foundation models, both in terms of the impact of genAI on society and on the further development of genAI tools. Our goals are threefold: to establish a shared vocabulary to facilitate collaboration and dialogue; to scope initial concerns to establish a framework for inquiry on which stakeholders can focus; and to shape future development of promising technologies proactively and positively.

The World Economic Forum's Global Future Council (GFC) on Data Equity[2] envisions this as a first step in a broader conversation, recognizing the need for further exploration and discussion to be comprehensively understood, scrutinised, and addressed. The issues are complex and interconnected. Tackling them now creates a unique opportunity to positively shape the future of these exciting, promising tools.

BOX 1 | **Definitions of key concepts**

To provide context and clarity, the following key concepts are highlighted:

– **Artificial intelligence** is a broad field that encompasses the ability of a machine or computer to emulate certain aspects of human intelligence for diverse tasks based on predetermined objectives.[3]

– **Machine learning** is a subset of artificial intelligence which utilizes algorithms to enable machines to identify and learn from patterns found in datasets.[4]

– **Generative AI** is a branch of machine learning that is capable of producing new text, images and other media, replicating patterns and relationships found in the training data.[5]

– **Foundation models** are a type of large-scale, machine-learning model that is trained on diverse multi-modal data at scale and can be adapted to many downstream tasks.[6]

– **Large language models** represent a subset of foundation models specializing in comprehending and generating human language, often employed for text-related functions. The latest iteration of LLMs facilitates natural conversations through advanced chatbot mechanisms.[7]

# 1 Classes of data equity

Effectively addressing the complexities of data equity mandates an appreciation of the diverse viewpoints held by various stakeholders regarding data. The academic literature has identified four distinct classes of data equity, which are closely interrelated:[8]

– **Representation equity** seeks to enhance the visibility of historically marginalized groups within datasets while also accounting for data relevancy for the target populations. The development of models primarily within the Global North introduces disparities in representation, potentially leading to systemic biases in subsequent decisions rooted in such data. A proactive approach is indispensable to ensure that AI training data and models authentically reflect all stakeholders without encoding biases.

– **Feature equity** seeks to ensure the accurate portrayal of individuals, groups and communities represented by data, necessitating the inclusion of attributes such as race, gender, location and income alongside other data. Without these attributes, it is often difficult to identify and address latent biases and inequalities.

– **Access equity** focuses on the equitable accessibility of data and tools across varying levels of expertise. Addressing transparency and visibility issues related to model construction and data sources is critical. Additionally, access equity also encompasses disparities in terms of AI literacy and the digital divide.

– **Outcome equity** pertains to impartiality and fairness in results. Beyond developing unbiased models, maintaining vigilance over unintended consequences that impact individuals or groups is necessary. Transparency, disclosure and shared responsibility are crucial to achieve fairness.

These four classes of data equity are particularly relevant to genAI, but not exhaustive. Two other prominent types of equity broadly applicable to technology that need to be considered are **procedural** and **decision-making equity**. These procedural elements underscore broad equity concerns and include transparent decision-making, fair treatment of workers who develop and deploy technology, and inclusive development and deployment practices.[9]

Going further, consideration must also be given to issues of temporal equity (sustainability and long-term impacts) and relational equity (fostering equitable stakeholder relationships). These latter issues are not unique to genAI or technology broadly and, as such, are beyond the scope of this paper. Nonetheless, they are acknowledged here as integral components of the overarching fabric of technology equity.

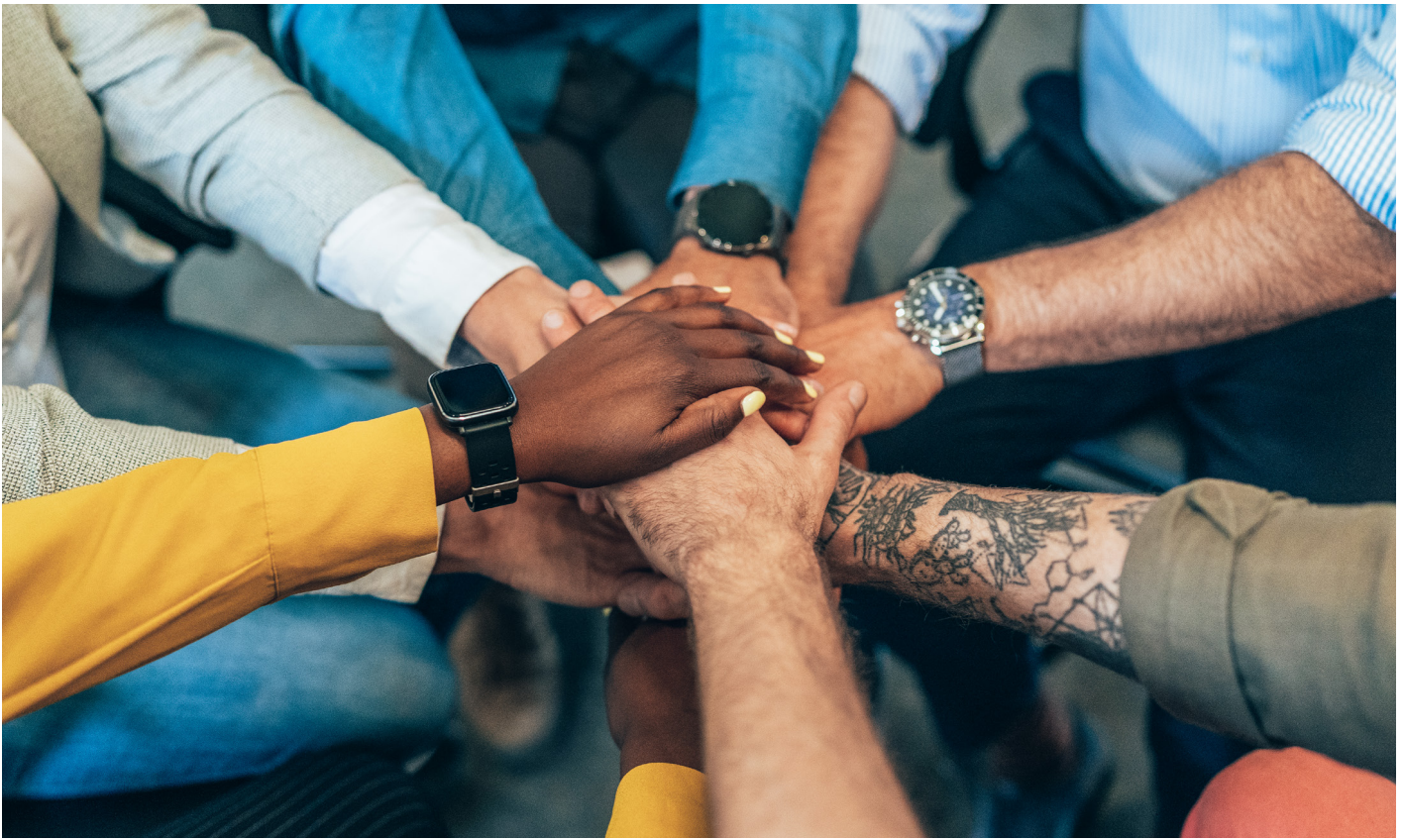

FIGURE 1 | Classes of data equity

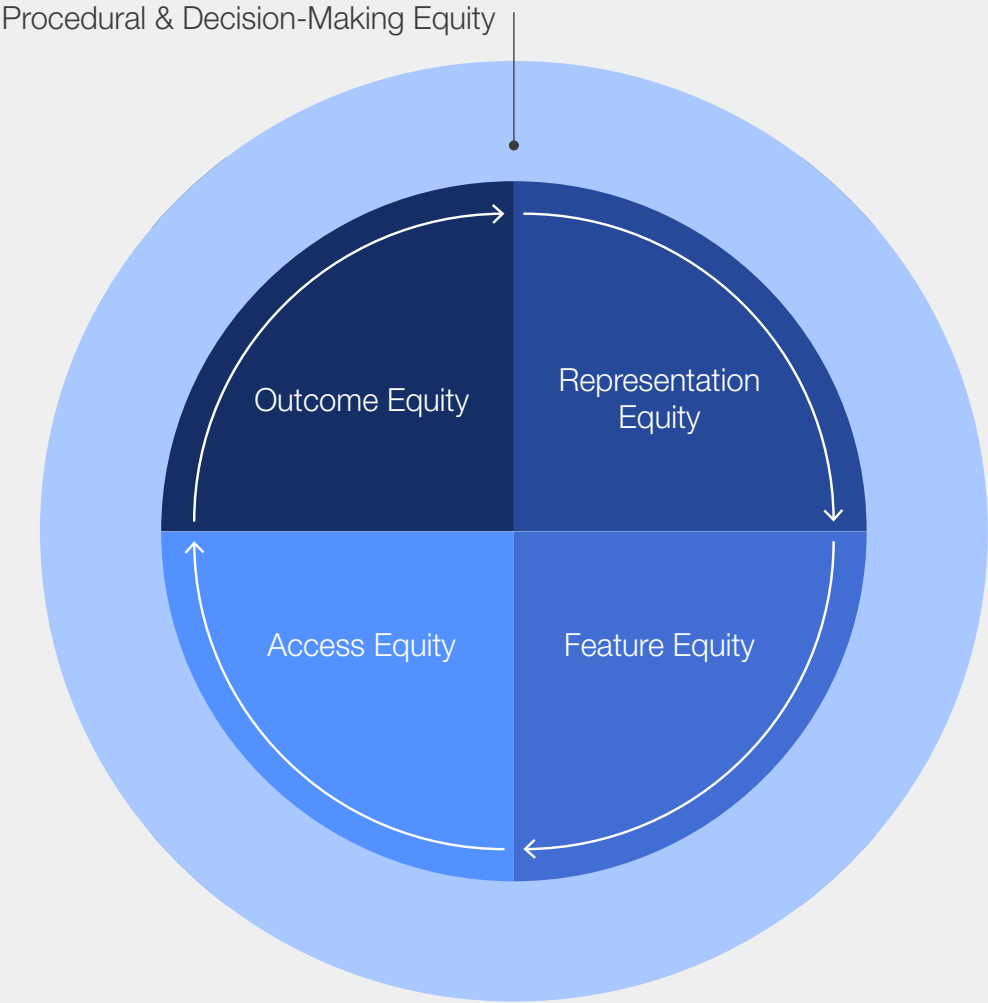

**Source:** World Economic Forum

**Figure 1:** The four classes of data equity issues are interconnected as well as influenced and impacted by equitable practices and considerations in procedures and decision-making.

# 2 | Data equity across the data lifecycle

A simplified representation is helpful in showing how data equity permeates the data lifecycle. At each stage, different classes of data equity raise specific challenges and concerns, illustrating the need for multifaceted approaches to mitigate potential harms.

FIGURE 2 | Data equity throughout the data lifecycle

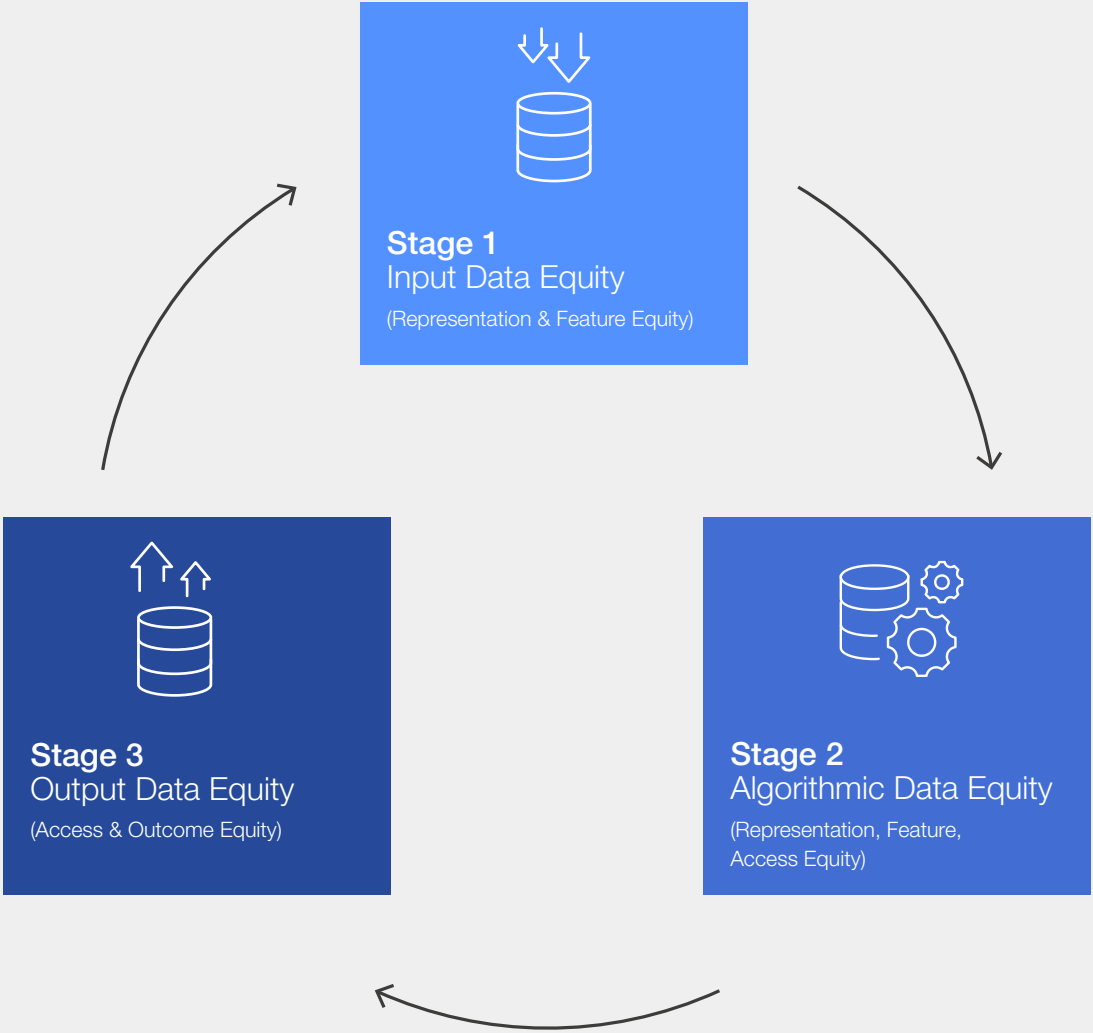

**Source:** World Economic Forum

**Figure 2:** Data equity across the data lifecycle. Ensuring data equity throughout the data lifecycle involves multiple stages: Stage 1 addresses the data that is used as input for developing foundation models. Stage 2 is the intermediary stage where algorithms are formulated and designed to analyse and interpret input data. Stage 3 focuses on the output data of genAI applications. Generated output may in some cases be used as input to further train foundation models, thereby exacerbating data equity challenges.

BOX 2 | **Data equity throughout the data lifecycle**

**Why focus on foundation models?**

Foundation models are at the core of many genAI tools. They are typically trained on large and complex datasets. Foundation models may encode results that reflect human prejudice, bias or misunderstanding; and training algorithms may discern incorrect relationships or context.

## Stage 1: Input data equity (representation and feature equity)

Input data equity centres on the data collected and used in building foundation models while also addressing the potential shortcomings this data might entail. As noted, foundation model training data may reflect societal inequities and result in societal bias. GenAI consequently generates outputs that mirror or amplify these patterns. Thus, ensuring equitable representation of diverse individuals, groups and communities in the datasets becomes pivotal to guarantee the relevance and accuracy of the generated outcomes.

This requirement extends beyond individual representation, encompassing the accurate portrayal of communities within information labelling. The promotion of fairness, bias mitigation and equal explanatory power practices is imperative for the outputs of foundation models to genuinely mirror the perspectives and realities of all individuals and groups inherent in the data. Moreover, the labels employed must be adaptable for use within algorithmic learning models.

Input data equity should also embrace the rights and well-being of data subjects. This encompasses aspects such as securing informed consent, just compensation for data contributors and annotators, and navigating the intricate trade-offs linked to data inclusion. These trade-offs are complex. While broader data inclusion may address equity concerns, it might concurrently escalate privacy worries through heightened surveillance. Similarly, generating new content can expand creative options but might not always ensure equitable compensation for the creators whose works contribute to the model's training.

The degree of anticipated data equity on the input side might vary based on the nature and objectives of the foundation models. Commercial applications, for instance, might prioritize transparency for end users, disclosing the scope and coverage of data, along with sensitivity analyses targeting specific groups. In other domains such as welfare allocation or legal applications, input side equity may demand the explicit inclusion of all pertinent communities to ensure genuine and tangible inclusivity.

## Stage 2: Algorithmic data equity (representation, feature, access equity)

Algorithmic data equity introduces a pivotal phase: the intermediary stage where algorithms are formulated and designed to interpret input data, thereby generating output results. This stage necessitates the incorporation of fairness, bias management and diversity inclusion in the algorithms' operations. It is imperative to ensure that these algorithms function as impartially as possible, refraining from perpetuating undesirable biases and accommodating diverse viewpoints. Attaining algorithmic data equity involves including a diverse array of perspectives in its design and assessing its influence on different demographic groups.

Algorithmic bias can emerge from several factors, such as the availability of suitable datasets. Concerns arise when culturally or geographically specific data is used to train models that will subsequently interact with populations not originally represented in the training data. For instance, models predominantly trained on North American or English-language content may struggle to offer accurate results for non-English-speaking populations or contexts outside the Global North.

Transparency also poses challenges as foundation models, which utilize neural networks, can produce complex and often opaque predictive outcomes. While other AI systems may allow for algorithmic transparency, the neural network-based learning process of genAI differs. Foundation models are pre-trained on vast datasets, which give them a broad base of knowledge. However, when fine-tuned or adapted to specific tasks, they initially rely on this general knowledge. As they are further trained on task-specific data, its predictions for that task can become more accurate, homing in on the intricate patterns and relationships within the new data they encounter.

This underscores the importance of exposing foundation models to diverse datasets, reflective of global communities. Moreover, fine-tuning algorithms to recognize the uniqueness of various regions and populations is vital to ensure the accurate understanding and prediction of relationships by foundation models, thus fostering balanced and equitable outcomes for users.

At the same time, given that digital literacy varies widely – and marginalized communities may be particularly underserved – ensuring global users' understanding of the models' capabilities and limitations becomes a significant equity concern for genAI's mass adoption.

## Stage 3: Output data equity (access and outcome equity)

Output data equity revolves around the fairness of tangible effects stemming from foundation model outputs. This encompasses benefits that directly arise from AI systems developed using this data. It involves asserting co-ownership rights over the AI system and advocating for the equitable sharing of benefits derived from the model.

Equitable distribution is also linked to the ability to share in the benefits generated by improvements to the AI system over time through iterative processes during the AI lifecycle. Instances where data collected in one region primarily bolsters the accuracy and performance of systems controlled by entities located in other regions underscore the importance of equitable sharing of these benefits with the originating communities.

Additionally, it is important for designers and implementers of AI systems to allocate resources to monitor and mitigate the disproportionate impacts on specific groups, reflecting biases and discrimination in the system's outputs, for example by making available remedial mechanisms. Data subjects and contributors have the right to influence the usage and governance of the AI system, particularly when it perpetuates harms or undesired effects. Similarly, those who contribute to the development of the system deserve to participate in the sharing of the profits or benefits generated by it.

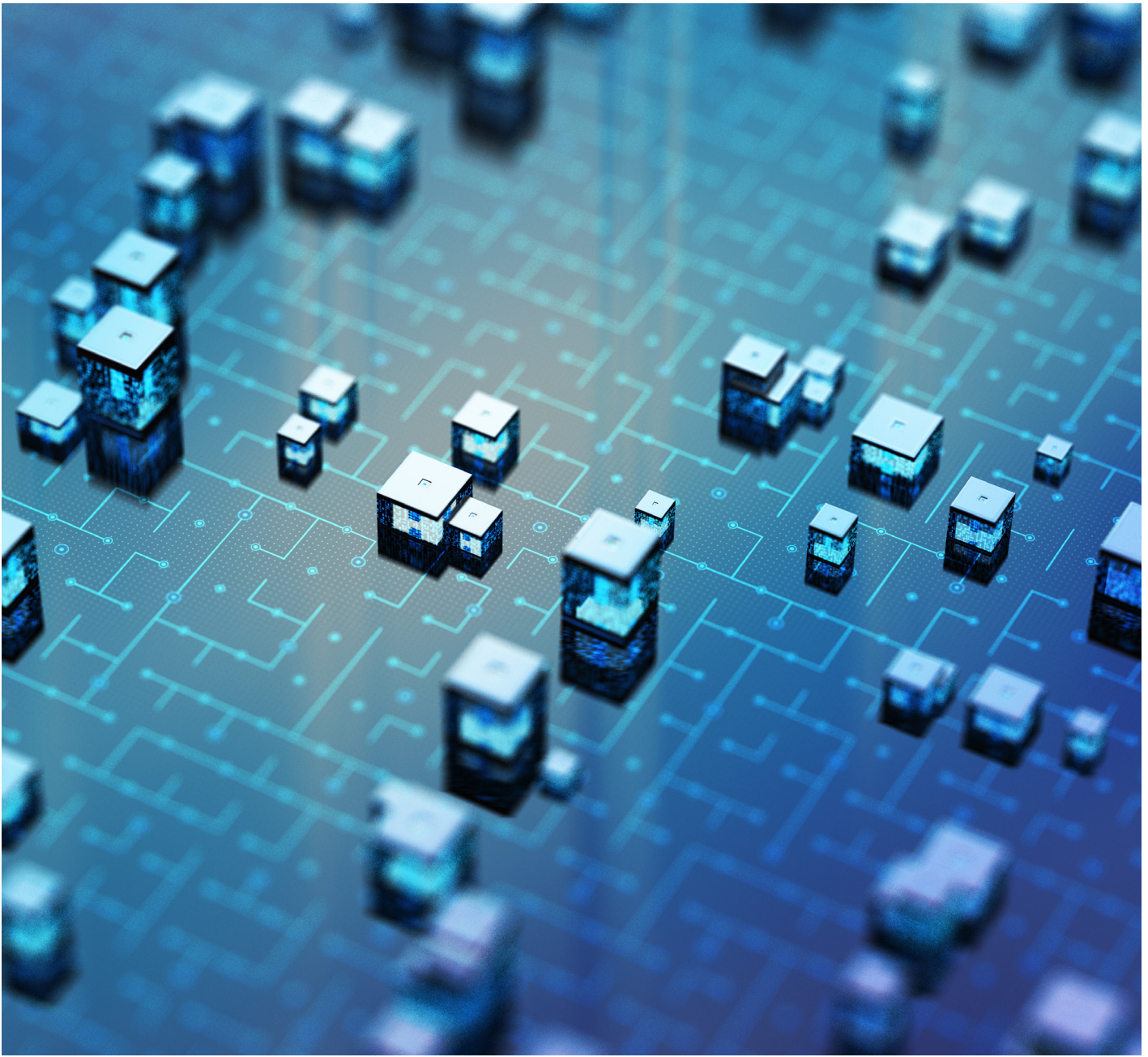

# 3 Data equity challenges in foundation models

The data equity challenges of foundation models in genAI are distinct from those in non-generative AI systems, highlighting a complex landscape that requires careful attention.

Training datasets requires innovative approaches to ensure accurate representation and consent. With genAI, the scale and diversity of such data raise other issues. Ethical dilemmas and privacy concerns arise from publicly available content, and the scale and ad hoc nature of data collection may render obtaining genuine consent impossible. Linguistic and cultural biases within training data, largely in English and from Western sources, can skew responses, favouring English-centric viewpoints and lead to an internationalization of dominant cultures. The release of genAI applications for mass consumption exacerbates automation bias, fuelled by insufficient transparency about model capabilities and limitations.

The unique features of foundation models – namely, the scale, volume and broad, often ambiguous, sourcing of data – complicate remediation. It is hard to pinpoint and correct specific data going into the model, which is further exacerbated by the ability of foundation models to generate entirely new content. This feature, while powerful for ongoing adaptation and learning, may further amplify bias and increase the difficulties associated with consent and Intellectual Property (IP) rights. Moreover, the datasets for foundation models are highly generalized and not built for specific use cases. A single foundation model may be used for multiple applications, those extending inequities across multiple domains or sectors.

Foundation models are continually learning and adapting. This unique feature creates further challenges given the scope, intricacy, size and training methods. As foundation models learn, algorithmic transparency, clarity and auditability become increasingly difficult. Secondly, reusing generated outputs can amplify existing biases. Recent research hints at the danger of "model collapse", in which a system seems to "forget" its initial data and worsens over time.[10] Moreover, given the size and complexity, replicating results or auditing models can be more challenging.

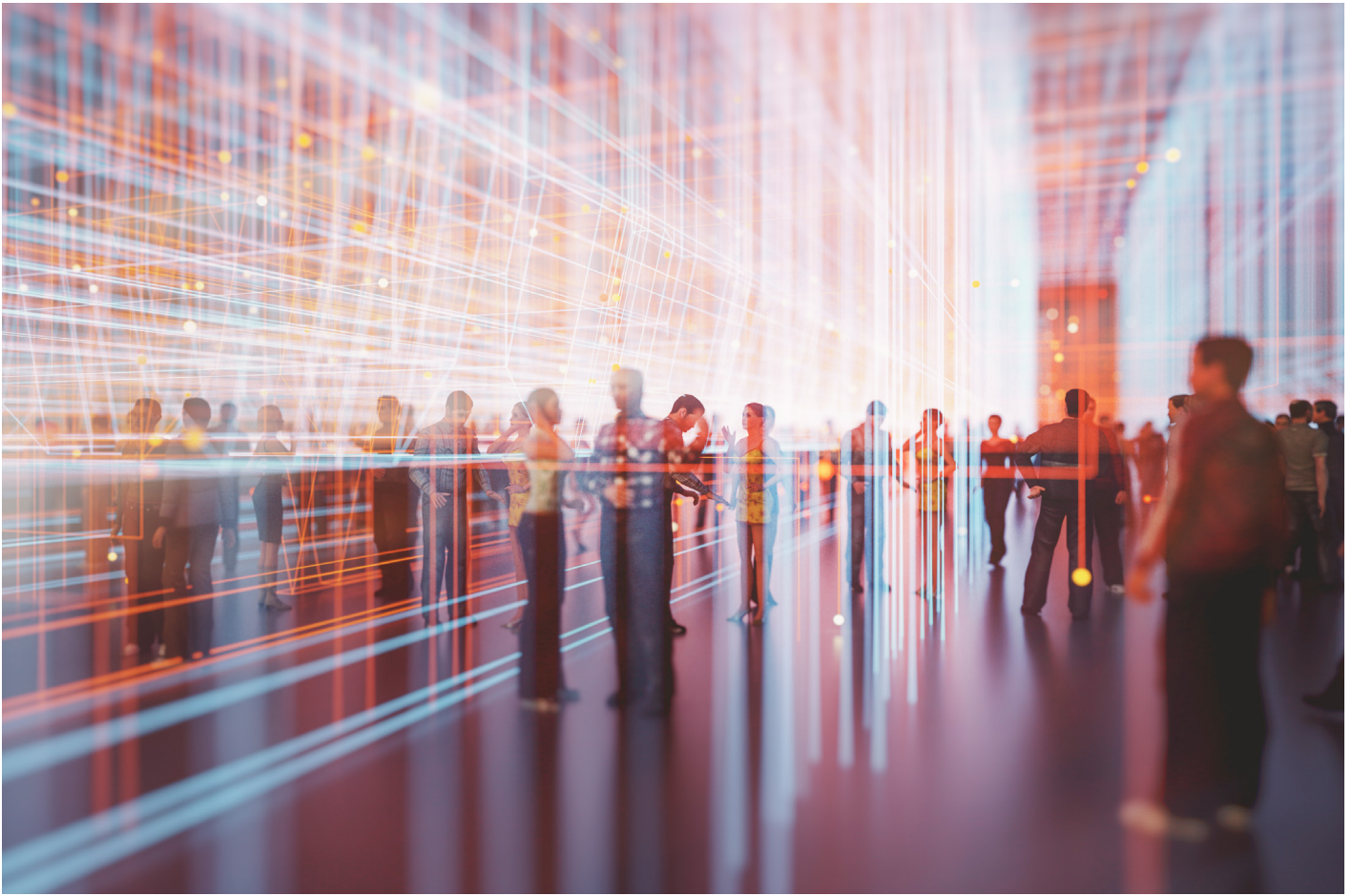

The table below summarizes some key differences between non-generative AI and generative AI's foundation models

TABLE 1 | Challenges: Non-Generative AI vs Generative AI

|  |  | Non-Generative AI | Foundation Models for GenAI |
|---|---|---|---|
| Unique Challenges | Scale of volume and source of data | Often uses smaller, curated datasets with known sources specifically relevant to the identified use case | – Uses massive datasets with often ambiguous, broad origins with no specific use case<br>– Hard to pinpoint and correct specific data |
| Unique Challenges | Generalizability vs specificity | Build for specific purpose(s) or task(s) | Designed for a broad range of tasks |
| Unique Challenges | Creation of novel content | Mostly analyses or predicts based on input data | – Can generate entirely new content, which may reflect or amplify biases explicit in the training data, or which may be misleading, inaccurate or false<br>– Generated content raises new consent and IP issues |
| Unique Challenges | Scale of impact | Because tools are developed for narrow use cases, impact is most relevant to the specific domain or application | A single model can have varied applications, thus extending or amplifying the effect of bias across multiple sectors and domains |
| Exacerbated Challenges | Opacity and complexity | Some models are interpretable | Scope, intricacy, size and training methods make algorithmic transparency and clarity especially challenging |
| Exacerbated Challenges | Feedback loops | Feedback loops might be less prevalent and controlled | – The continuous refinement process of standard training methods can reinforce biases<br>– Reusing generated outputs can amplify existing bias |
| Exacerbated Challenges | Reproducibility and accountability | Easier to reproduce and pinpoint source of biases | Due to the size and complexity, replicated results or auditing can be more challenging |
| Exacerbated Challenges | Internationalization of dominant culture | Since the model is domain-specific, the risk of spreading a dominant culture is lesser | Broad applicability risks minimizing or overlooking the needs of specific communities. This can inadvertently promote dominant cultural viewpoints globally |

**Table 1:** Unique and exacerbated challenges in the case of non-generative AI versus foundation models for generative AI. It is important to note that this is a non-exhaustive list.

# 4 Focus areas for key stakeholders

Addressing data equity is a complex undertaking and will require the active, engaged participation of many individuals, groups and communities. As a starting point, we propose various pathways and actions stakeholders should take to ensure data equity when interacting with foundation models. Three major groups of stakeholders can be distinguished:

– **Those that are responsible for driving and governing the societal use of AI**: AI-creating, AI-using organizations and policy-makers.

– **Those that are impacted by or are the end users of AI systems**: The public and communities. When it comes to the public and communities as stakeholders, there is an inherent power asymmetry between them and the other stakeholders due to differences in both capacities to use AI and levels of data literacy. It is important that those accountable for driving and governing the societal use of AI ensure meaningful engagement with the public and communities.

– **Those that can bridge concerns between the accountable stakeholders and the public and communities**: Civil society, with a focus on capacity building and developing representation for the public and communities with organizations that are responsible for AI.

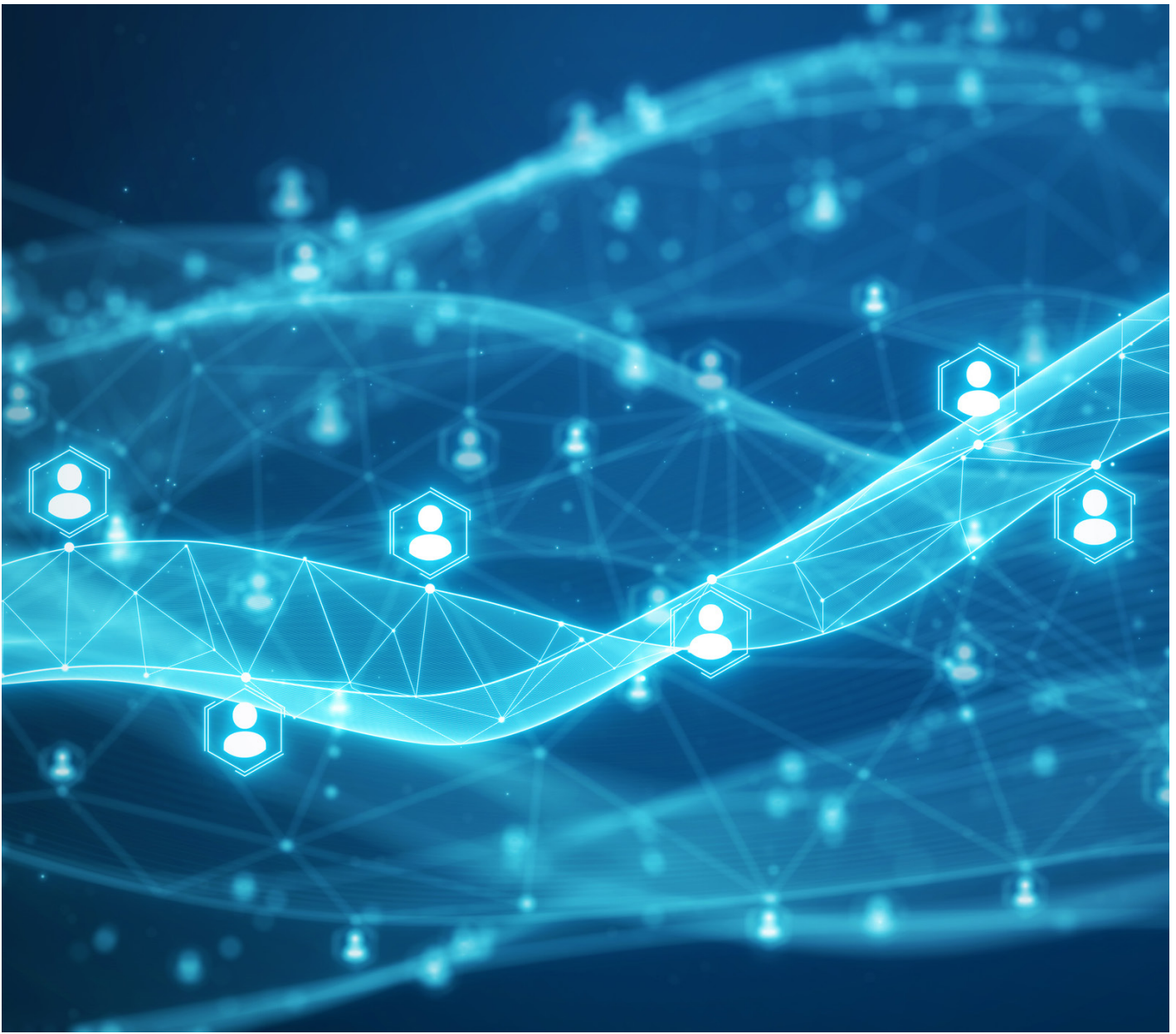

TABLE 2 | Focus areas, potential outcomes and example pathways for key stakeholders to ensure data equity in foundation models.

## Those responsible for driving and governing societal use of AI

| Stakeholder | AI-creating organizations | AI-using organizations | Policy-makers and regulators |
|---|---|---|---|
| **Focus areas** | – Data collection and labelling<br>– Data privacy and security<br>– Transparency, traceability, and explainability<br>– Mitigation strategies (incl. fairness and bias mitigation)<br>– Continuous model evaluation<br>– Inclusive model design | – Responsible AI practices<br>– Data access and usage, incl. data privacy and security<br>– Disclosure to impacted communities<br>– Continuous monitoring<br>– Mitigation strategies (incl. fairness and bias mitigation)<br>– Context appropriate AI-human decision-making balance | – Develop ethical guidelines and standards[10]<br>– Develop regulatory frameworks, including audits<br>– Consideration of public interest<br>– AI risk classifications<br>– Clear delineation of rights of data subjects and contributors regarding AI<br>– Raise public awareness |
| **Potential outcomes** | – Meaningful transparency<br>– Model traceability for better quality control<br>– Effective accountability, incl. clear pathways for accountability (both external and internal)<br>– Implementation of assessment measures<br>– Facilitate continuous independent audits<br>– Collaborate with content generators | – Public disclosure of AI system usage<br>– Implement responsible AI governance frameworks<br>– Adopt standard practices<br>– Develop clear methodologies<br>– Ensure clear guidelines of automation circuit-breakers | – Establish standards and enact regulation<br>– Human-rights based approach<br>– Universal AI ethics<br>– Set an observatory body to ensure regulatory engagement and enforcement[11]<br>– Engage multistakeholder community, incl. industry, academia, civil society, and public<br>– Including meaningful engagement with stakeholders from the Global South |
| **Example pathways** | – Open-source a representative portion of data<br>– Pre-launch and continuous auditing and monitoring of model behaviour<br>– Create and use public feedback channels<br>– Build tools that provide greater transparency | – Due diligence prior to deployment<br>– Create and utilize public feedback channels<br>– Ethical guidelines and training | – Consult global AI experts<br>– Facilitate regulatory sandboxes as a best practice to design and test genAI systems<br>– Educate judiciary<br>– Implementation of Indigenous data sovereignty frameworks[12] |

## Those using and impacted by AI systems

| Stakeholder | | Civil society groups | Public | Communities |
|---|---|---|---|---|
| | **Focus areas** | – Bridge gap between AI organizations and public by raising awareness through advocacy efforts<br>– Promote ethical practices | – Increased awareness of AI<br>– Understanding of AI ethics<br>– Engagement with AI stakeholders | – Impact of AI on affected communities<br>– Participation in AI decision-making discussions |
| | **Potential outcomes** | – Develop accessible research and awareness material for the general public<br>– Develop ethical practice codes and model legislation | – Greater public awareness on how AI might influence issues and topics the public cares about<br>– Engage with stakeholders in public debates | – Understand impact of AI on everyday life<br>– Actively participate in advocacy campaigns<br>– Capacity-building for those using AI |
| | **Example pathways** | – Public awareness campaigns<br>– Create data equity toolkits and resources | – Become educated on AI<br>– Learn about and participate in advocacy campaigns<br>– Hold stakeholders accountable | – Report and share observations with policy-makers<br>– Consider what data equity means in specific communities, such as in the case of Indigenous data sovereignty |

**Note:** Academia can either be part of AI-creating organizations, civil society, or communities, depending on the focus areas and the research undertaken.

# 5 Discussion

This paper has introduced main ideas and concepts about data equity. It is important to recognize, however, that data equity will have sector-specific considerations across all stages of the data cycle discussed above (input, algorithmic, output).

Addressing data equity in the use of foundation models (and AI in general) requires greater transparency about the limitations, capabilities and therefore the application of data to AI in different contexts. As AI is being used to inform decision-making, it highlights the need to consider the human dimensions and socio-technical elements of both the development and utilisation of AI. Acknowledgment of such limitations and the required correctives may be informed by the nature of the data used, the kind of AI model and the sensitivity of the application space.

As digital society evolves, genAI application functions will increasingly become intelligently autonomous to an even greater extent. AI is expected to be widely available at an industrial scale in all sectors and become less expensive, more convenient and more easily accessible to use. This widespread availability lends itself to a general tendency to overuse genAI models. A key problematic result of this would be encoding data inequities, thereby perpetuating epistemic inequities.[14] It is thus also critical to evaluate the utility of genAI for a given use case; in some scenarios, more traditional data science or AI approaches might be more relevant and useful. Keeping an appropriate AI vs "human decision-making" balance in different contexts reduces the chances of perpetuating these inequities when foundation models are used.

At the same time, it is also important to recognize the potential of generative AI in enhancing data equity. GenAI applications may be used for example to improve data analysis, provide further explanation and increase access to data. For this briefing paper, we decided to focus specifically on the challenges of data equity in generative AI, given the importance of addressing these challenges early on in the adoption of genAI applications. As a result, the opportunities of genAI for data equity fall outside the purview of this briefing paper.

# Conclusion

GenAI promises immense potential to drive digital and social innovation, including improving efficiency, enhancing creativity and augmenting existing data. Generative AI has the potential to democratize access and usage of technologies, thereby bridging the digital divide.[15] However, if left unchecked, it could further engrain inequities.

As these systems rapidly advance, only a small window exists to act decisively. It is crucial to integrate data equity and ethical considerations into every phase of genAI's development, from dataset collection to model training and model output. Ignoring issues at this moment will only amplify the inequities and increase the data and digital divides in societies. Now is the time to create definitional terms for collaboration in order to develop methods and processes that can be incorporated into technological development. While data equity concepts have existed in systems and methods for some time, the rise of genAI marks an urgent moment to foster dialogue and collaborative efforts across all sectors of society.

This briefing paper represents a first step in exploring and promoting data equity in the context of genAI. The proposed definitions, framework and recommendations are intended to be applicable to proactively and positively shape the future development of promising genAI technologies.

Through this and future work, the World Economic Forum's Global Future Council on Data Equity seeks to ensure equitable results throughout the broader digital economy, enabling fair and widespread global sharing of societal outcomes and benefits, and to start a dialogue on data equity among all stakeholders.

It is only by identifying and acknowledging different types of systemic inequities that we can address them and work towards more comprehensive and inclusive solutions, to ensure shared benefits of generative AI. We look forward to continuing the conversation and working towards enhanced data equity.

# Contributors

Global Future Council on Data Equity 2023-2024

The World Economic Forum's network of Global Future Councils is the world's foremost multistakeholder and interdisciplinary knowledge network dedicated to promoting innovative thinking to shape a more resilient, inclusive and sustainable future.

## Global Future Council on Data Equity Council Members

**JoAnn Stonier (co-chair)**
Mastercard Fellow, Data & AI, Mastercard

**Lauren Woodman (co-chair)**
Chief Executive Officer, DataKind

**Majed Alshammari**
Special Adviser, Data Governance, Saudi Data and AI Authority (SDAIA)

**Renée Cummings**
Data Science Professor & Data Activist in Residence, University of Virginia

**Nighat Dad**
Founder and Executive Director, Digital Rights Foundation

**Arti Garg**
AI Chief Strategist, Hewlett Packard Enterprise

**Alberto Giovanni Busetto**
Group Senior Vice-President; Head, Data and Artificial Intelligence, Adecco Group

**Katherine Hsiao**
Executive Vice-President; Head, Health and Life Sciences, Palantir Technologies

**Maui Hudson**
Associate Professor and Director, Te Kotahi Research Institute, University of Waikato

**Parminder Jeet Singh**
Digital Society Researcher

**David Kanamugire**
Chief Executive Officer, National Cyber Security Agency of Rwanda

**Astha Kapoor**
Co-Founder, Aapti Institute

**Zheng Lei**
Professor, Fudan University

**Jacqueline Lu**
President and Co-Founder, Helpful Places

**Emna Mizouni**
Chief Executive Officer, Digital Citizenship

**Angela Oduor Lungati**
Executive Director, Ushahidi

**María Paz Canales Loebel**
Head of Legal, Policy and Research, Global Partners Digital

**Arathi Sethumadhavan**
User Research Scientist, Technology and Society, Google

**Sarah Telford**
Lead, Centre for Humanitarian Data, United Nations Office for the Coordination of Humanitarian Affairs (OCHA)

## World Economic Forum

**Supheakmungkol Sarin**
Head of Data and Artificial Intelligence Ecosystems, Centre for the Fourth Industrial Revolution; Council Manager, Global Future Council on the Future of Data Equity

**Kimmy Bettinger**
Lead, Expert and Knowledge Communities, Centre for the Fourth Industrial Revolution

**Stephanie Teeuwen**
Early Careers Programme – Data Policy, Centre for the Fourth Industrial Revolution

# Acknowledgements

**Talal Altook**
Fellow, Artificial Intelligence and Machine Learning, Centre for the Fourth Industrial Revolution, World Economic Forum

**Genta Ando**
Fellow, AI Governance Alliance, Centre for the Fourth Industrial Revolution, World Economic Forum

**Jos Berens**
Data Policy Officer, Centre for Humanitarian Data, United Nations Office for the Coordination of Humanitarian Affairs (OCHA)

**Sebastian Buckup**
Head of Network and Partnerships, Centre for the Fourth Industrial Revolution, World Economic Forum

**John Bradley**
Lead, Metaverse, Centre for the Fourth Industrial Revolution, World Economic Forum

**Kasia Chmielinski**
Principal, Data Nutrition Project

**Tenzin Chomphel**
Coordinator, Data Policy, Centre for the Fourth Industrial Revolution, World Economic Forum

**Daisuke Fukui**
Fellow, Advancing Cross-Border Data Flows, Centre for the Fourth Industrial Revolution, World Economic Forum

**Devendra Jain**
Lead, Digital Transformation, Centre for the Fourth Industrial Revolution, World Economic Forum

**Benjamin Larsen**
Lead, Artificial Intelligence and Machine Learning, Centre for the Fourth Industrial Revolution, World Economic Forum

**Cathy Li**
Head of AI, Data and Metaverse, Centre for the Fourth Industrial Revolution, World Economic Forum

**Sandra Waliczek**
Centre Curator, Blockchain and Digital Assets, World Economic Forum

**Karla Yee Amezaga**
Lead, Data Policy, Centre for the Fourth Industrial Revolution, World Economic Forum

# Production

# Endnotes

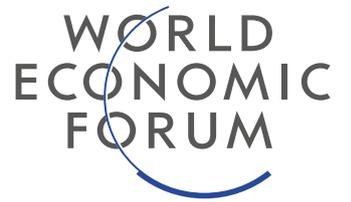

COMMITTED TO
IMPROVING THE STATE
OF THE WORLD

The World Economic Forum,
committed to improving
the state of the world, is the
International Organization for
Public-Private Cooperation.

The Forum engages the
foremost political, business
and other leaders of society
to shape global, regional
and industry agendas.

World Economic Forum
91–93 route de la Capite
CH-1223 Cologny/Geneva
Switzerland

Tel.: +41 (0) 22 869 1212
Fax: +41 (0) 22 786 2744
contact@weforum.org
www.weforum.org